\documentclass{iaus}
\usepackage{graphicx}

\title[Galaxy Mergers and Interactions at High Redshift] %% give here short title %%
{Galaxy Mergers and Interactions at High Redshift}

\author[Christopher J. Conselice]   %% give here short author list %%
{Christopher J. Conselice%
\affiliation{ School of Physics \& Astronomy, University of Nottingham, NG7 2RD, England \\}}

\pubyear{2006}
\volume{235}  %% insert here IAU Symposium No.
\pagerange{1-4}
\date{?? and in revised form ??}
\jname{Galaxy Evolution Across the Hubble Time }
\editors{F. Combes and J. Palous, eds.}

\def\solm{M$_{\odot}\,$}

\begin{document}
\maketitle

\begin{abstract}

In this review we discuss the evidence for galaxy interactions and mergers in 
the distant universe and the role of mergers in 
forming galaxies.  Observations show that the fraction of massive ($M> M_{*}$)
galaxies involved in major mergers is roughly 5-10\% at $z \sim 1$. The
merger fraction however increases steeply for the most 
massive galaxies up to $z \sim 3$, where the merger fraction is 
50$\pm$20\%. Using N-body models of the galaxy merger process at a variety 
of merger conditions, merger mass ratios, and viewing angles this merger 
fraction can be converted into a merger rate, and mass accretion rate due 
to mergers. A simple integration of the merger rate shows that a typical
massive galaxy at $z \sim 3$ will undergo 4-5 major mergers between
$z \sim 3$ and $z \sim 0$,
with most of this activity, and resulting mass assembly, occurring at 
$z > 1.5$.

\end{abstract}

\firstsection % if your document starts with a section,
\section{Introduction}

Do galaxies merge? Does the formation of galaxies rely on merging?  The
answer to the first question is undoubtedly yes.  There are undeniable and
famous ongoing galaxy mergers in the nearby universe, such as the
Antennae, as well as accretion of galaxy satellites into our own galaxy.
Perhaps mergers are rare however, and perhaps the low level merging
seen in our own galaxy is a minor contributor to its mass and evolution.

On the other hand, merging may be the dominate method whereby galaxies
acquire their mass, and this activity may drive the evolution we see in distant
galaxies (e.g., Hopkins et al. 2006).  It is furthermore possible
that galaxy merging is not only the driving force behind galaxy
formation, but may also induce star formation and black hole growth, and
has been conjectured to be the link between galaxies and quasars.
We are just beginning to explore observationally the idea  that merging is the
dominate method whereby galaxies form, and understanding its role
in a cosmological context will likely be one of the major debates in 
extragalactic astrophysics in the coming years.

There are many reasons to expect that galaxies form through mergers. Perhaps 
the most overwhelming is that the largely accepted cosmological
model -- a $\Lambda$ dominated Cold Dark Matter (CDM) based universe -- 
explicitly
predicts that galaxies should form in the merger process.  If mergers are
not responsible for the majority of massive galaxy formation, it would
require a reevaluation of our assumptions concerning dark matter and
the nature of baryonic physics.  More likely, observational studies
of the merger history will allow us to test galaxy formation models
on a very fundamental level. 

Despite the importance of understanding the history of galaxy
merging, there is very little known about its history, and the
available results still contain large uncertainties.  Despite this, it is worth
reviewing the evidence for the role of mergers
in the formation of galaxies.  We argue that the available evidence 
suggests that mergers are the dominate process in forming at least the
most massive galaxies in the universe. 

\section{Observed Merger History}

\subsection{Merger Fractions}

\begin{figure}
\includegraphics[height=2.5in,width=5.5in,angle=0]{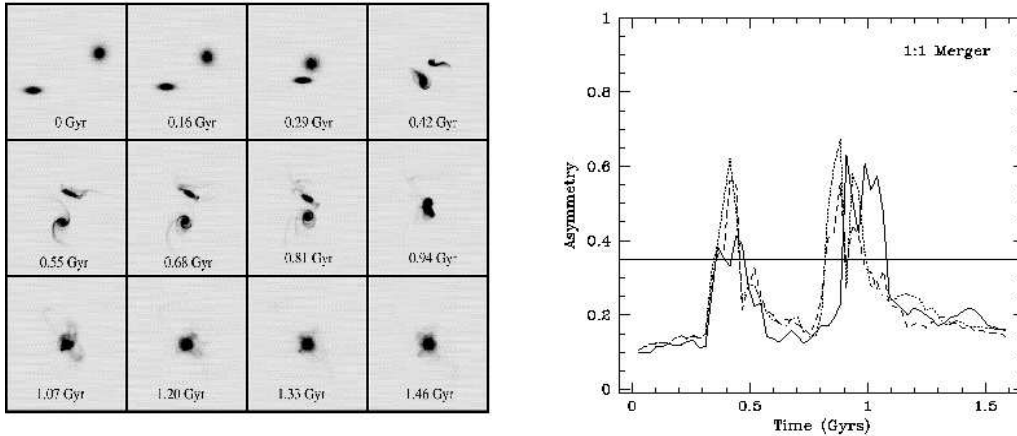}
\vspace{-1cm}
  \caption{N-body model of two disk galaxies with the same
mass merging.  The number on the bottom of each simulated
image shows the time in the simulation. The resulting measured
asymmetry for each galaxy is shown on the right hand panel 
(Conselice 2006).}\label{fig:contour}
\end{figure}

The measurements of the history of galaxy merging is sometimes thought to be
ambiguous. However this often results from comparing measured galaxy merger
fractions (f$_{\rm gm}$) and merger rates ($\Re$) using different techniques
to identify the total number of galaxies involved in mergers (N$_{\rm gm}$) at 
various wavelengths ($\lambda$), as well as at different luminosity 
(M$_{\rm B}$), and stellar mass (M$_{*}$) ranges.  Alter any 
of these properties, and the measured merger fraction will change. The
observed merger fraction is the number of
galaxies determined to be a merger by a technique within a given
set of observables divided by the
total number of galaxies within that observable range $({\rm N_{T}})$: $${\rm f_{gm}} ({\rm M_{*},M_{B}},z,\lambda) = \frac{{\rm N_{gm}}  ({\rm M_{*},M_{B}},z,\lambda)}{{\rm N_{T}} ({\rm M_{*},M_{B}},z,\lambda)}.$$

\noindent Very different
merger fractions for the same population of galaxies can be uncovered by using
techniques that differ in merger sensitivity.  Normalising the merger fraction
by its time-scale sensitivity gives the merger rate, and normalising the
merger rate by the mass sensitivity gives the mass accretion rate. The
merger rate and the mass accretion rate are the preferred
measured quantities. These quantities are hard to measure, but early 
attempts to measure them are promising (Conselice 2006).

There are only a few methods for measuring the merger
fraction, all of which have their limitations and advantages.  The oldest
and perhaps most straightforward observational method is to
look for galaxies in pairs (e.g.,  Lin et al. 2004). However, pair studies 
require spectroscopy of complete samples of field galaxies, and
have only successfully been applied up to $z \sim 1.4$.  The pair
fraction, and resulting merger fraction based on pairs of similar
luminosities, increases slightly up to $z \sim 1$ (see Bridge et al.
2006).  The other method, particularly useful at $z > 1$, is to 
look for galaxies that
are peculiar in their morphology or kinematics, and to use
this as a measure of the merger fraction (Conselice et al. 2003;
Lavery et al. 2004; Papovich et al. 2005; Conselice et al. 2005; 
Lotz et al. 2006).  For example, studies utilising the asymmetry
index (Conselice et al. 2000; Conselice 2003) show that
the merger fraction is very high at $z \sim 2.5$, with the 
M$_{*} > 10^{10}$~\solm merger fraction at $\sim 50$\%, and 
steeply declining as (1+z)$^{3\pm0.3}$.

Determining the history of galaxy interactions, as opposed to
active mergers, is more difficult.  In some sense, looking for
galaxies in pairs (Patton et al. 2002) gives us a good idea for
the likely interaction history, as close galaxies will be interacting
without necessarily merging.   While some increase in the star
formation rate is seen for galaxies in close pairs (e.g., Barton et al.
2003), it is not clear what the global role interactions play in
the increase of stellar mass in galaxies.  Recent investigations
suggest that the merger and interaction history together contribute 
a large fraction of the star formation at $z < 1$ (Bridge et al. 2006).

\subsection{Galaxy Merger Rates}

The measurements of galaxy merger rates, and mass accretion rates
due to the merger process, are the ultimate goal of galaxy merger
studies. Observationally, it is not trivial to determine the
merger rate, and very often it is necessary to utilise models
in some regard. The basic problem is understanding how
long during a merger your particular method of finding mergers is
sensitive.  This quantity is in principle easy to measure for
systems in pairs using dynamical friction arguments; yet it is
impossible to constrain the relative velocities between galaxy pairs
in the tangential direction or know the spatial separation in the
radial direction.  Considering all of these uncertainties the
time-scale for two galaxies at roughly 20 kpc separation will merge
within 0.5-1 Gyr.

\begin{figure}
\includegraphics[height=2.5in,width=5.3in,angle=0]{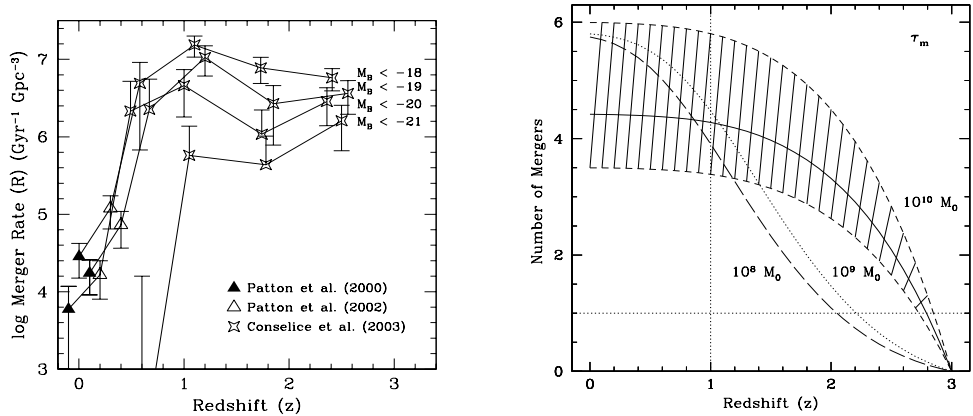}
  \caption{Evolution of the merger rate, in units of Gyr and co-moving
Gpc$^{3}$,
as a function of redshift and observed magnitude (left panel), and the 
empirically determined integrated number of major mergers since 
$z \sim 3$.  These merger rates and histories are taken from
merger fraction data in Conselice et al. (2003) and Patton
et al. (2002, 2000).}\label{fig:contour}
\end{figure}

Converting a measured merger fraction derived morphologically has
its own advantages and problems. In principle it could be easier
to measure the time-scale for merging sensitivity.  Analysing a set
of N-body simulations of mergers, Conselice (2006) determine the
first time-scales on the merger process using the CAS methodology 
(Conselice 2003)
for finding mergers.  An example of this is shown in Figure~1 where
the morphological evolution of two disk galaxies merging is shown,
as well as the resulting asymmetry computation for this simulation.
Based on these simulations, viewed at various angles and including
galaxies in various orbital configurations, a time-scale in
which a merging galaxy would be found within the CAS system
can be derived. For galaxies with masses 10$^{11}$ \solm the merger
sensitivity  is $\tau_{\rm m} = 0.38\pm0.1$ Gyr. Using this
merger time-scale we can calculate the merger rate
evolution for galaxies up to $z \sim 3$ (Figure~2).

A new, and potentially powerful, method for measuring the evolution of
the merger rate is to determine how the correlation function of
galaxies changes over time (e.g., Masjedi et al. 2006).  
The inferred merger rate since $z \sim 0.36$ for luminous elliptical
galaxies show that mergers since at least this time have been rare.  
A similar method applied up to $z \sim 1$
demonstrates some evolution, perhaps one merger per massive galaxy
since $z \sim 1$ (Bell et al. 2006).

\section{Implications of the Merger History}

We can use the merger rate to determine the total number of
mergers an average galaxy undergoes since the beginning of our
observational epoch. By integrating the merger rate
since $z \sim 3$ we find that a typical massive
galaxy with M$_{*} > 10^{10}$ \solm undergoes 
4.4$_{-0.9}^{+1.6}$ mergers (Figure~2). An additional feature of 
the N-body models 
analysed in Conselice (2006) is the ability to determine the merging
galaxy mass ratios that can produce high asymmetries.  
The result of this is that the CAS method is only
sensitive to major mergers, that is mergers with a mass
ratio of 1:3 or lower (see also Hernandez-Toledo
et al. 2005).   This also allows us to determine how much
mass is likely added to galaxies due to the merger process
since $z \sim 3$. The result is that a galaxy which undergoes
on average 4.4 major mergers can increase its total mass by
a factor of $\sim10$.  Due to the advent of deep X-ray imaging,
we can also now test the idea that galaxy interactions and
black hole build up are related. The evidence for
a connection is ambiguous (Grogin et al. 2005; Pierce et al.
2006), unless there is a significant delay between the merger
and AGN activity.

The uncertainties on merger fraction histories and rates
are still uncomfortably large. Future surveys utilising high-resolution
wide-field near-infrared imaging will revolutionise this field.
Complete redshift surveys at $z > 1.5$ will furthermore allow
us to determine the pair and interaction history at higher redshifts,
although such a survey will be largely impossible until the existence
of 20-30m telescopes. There are also largely no constraints on the
minor merger history, although methods to
measure these are in development. These and
other technical advances in the coming years will advance
our knowledge of the role of mergers and interactions in galaxy
formation considerably.

\end{document}